\documentclass[twocolumn,traditabstract]{aa}  

\usepackage{amsmath}
\usepackage{fixltx2e}
\usepackage[english]{babel}
\usepackage{graphicx}
\usepackage{epstopdf}
\usepackage{epsf,color}
\usepackage[mathscr]{eucal}
\usepackage{amsmath}
\usepackage{amssymb,amsfonts}
\usepackage{natbib}
\usepackage{graphicx}
\usepackage{txfonts}
\usepackage{dsfont}
\definecolor{Mygreen}{rgb}{0.75, 0.0, 0.0}
\definecolor{Mypink}{rgb}{1.0, 0.0, 0.5}
\definecolor{Myred}{rgb}{0.7, 0.0, 0.0}
\usepackage{hyperref}
\usepackage[vertfit]{breakurl}
\usepackage{float} 
\usepackage{color}

\bibpunct{(}{)}{;}{a}{}{,}
\newfont{\gwpfont}{cmssq8 scaled 1000}
\mathchardef\mhyphen="2D

\newcommand {\apgt} {\ {\raise-.5ex\hbox{$\buildrel>\over\sim$}}\ }
\newcommand {\aplt} {\ {\raise-.5ex\hbox{$\buildrel<\over\sim$}}\ }

\newcommand{\asec}{$^{\prime \prime}$}
\newcommand{\asecs}{$^{\prime \prime}\ $}
\newcommand{\amin}{$^{\prime}$}
\newcommand{\amins}{$^{\prime}\ $}

\defcitealias{arnaud2010}{A10}
\defcitealias{cavagnolo2009}{C09}
\defcitealias{bulbul2010}{B10}
\defcitealias{vikhlinin2006a}{V06}

\begin{document}

\title{A multi-instrument non-parametric reconstruction of the electron pressure profile in the
  galaxy cluster CLJ1226.9+3332}
\author{C.~Romero\inst{\ref{IRAMF}} \thanks{Corresponding author: Charles Romero, \url{romero@iram.fr}}			        
\and M.~McWilliam\inst{\ref{LPSC},\ref{ICL}}
\and J.-F.~Mac\'ias-P\'erez\inst{\ref{LPSC}}	      
\and  R.~Adam \inst{\ref{LPSC},\ref{OCA}}
\and  P.~Ade \inst{\ref{Cardiff}}
\and  P.~Andr\'e \inst{\ref{CEA}}
\and  H.~Aussel \inst{\ref{CEA}}
\and  A.~Beelen \inst{\ref{IAS}}
\and  A.~Beno\^it \inst{\ref{Neel}}
\and  A.~Bideaud \inst{\ref{Neel}}
\and  N.~Billot \inst{\ref{IRAME}}
\and  O.~Bourrion \inst{\ref{LPSC}}
\and  M.~Calvo \inst{\ref{Neel}}
\and  A.~Catalano \inst{\ref{LPSC}}
\and  G.~Coiffard \inst{\ref{IRAMF}}
\and  B.~Comis \inst{\ref{LPSC}}
\and  F.-X.~D\'esert \inst{\ref{IPAG}}
\and  S.~Doyle \inst{\ref{Cardiff}}
\and  J.~Goupy \inst{\ref{Neel}}
\and  C.~Kramer \inst{\ref{IRAME}}
\and  G.~Lagache \inst{\ref{LAM}}
\and  S.~Leclercq \inst{\ref{IRAMF}}
\and  J.-F.~Lestrade \inst{\ref{LERMA}}
\and  P.~Mauskopf \inst{\ref{Cardiff},\ref{Arizona}}
\and  F.~Mayet \inst{\ref{LPSC}}
\and  A.~Monfardini \inst{\ref{Neel}}
\and  E.~Pascale \inst{\ref{Cardiff}}
\and  L.~Perotto \inst{\ref{LPSC}}
\and  G.~Pisano \inst{\ref{Cardiff}}
\and  N.~Ponthieu \inst{\ref{IPAG}}
\and  V.~Rev\'eret \inst{\ref{CEA}}
\and  A.~Ritacco \inst{\ref{IRAME}}
\and  H.~Roussel \inst{\ref{IAP}}
\and  F.~Ruppin \inst{\ref{LPSC}}
\and  K.~Schuster \inst{\ref{IRAMF}}
\and  A.~Sievers \inst{\ref{IRAME}}
\and  S.~Triqueneaux \inst{\ref{Neel}}
\and  C.~Tucker \inst{\ref{Cardiff}}
\and  R.~Zylka \inst{\ref{IRAMF}}}

\institute{
Institut de RadioAstronomie Millim\'etrique (IRAM), Grenoble, France
  \label{IRAMF}
\and
  Laboratoire de Physique Subatomique et de Cosmologie, Universit\'e Grenoble Alpes, CNRS/IN2P3, 53, avenue des Martyrs, Grenoble, France
  \label{LPSC}
  \and
  Imperial College London, Kensington, London SW7 2AZ, UK
  \label{ICL}
\and  
  Laboratoire Lagrange, Universit\'e C\^ote d'Azur, Observatoire de la C\^ote d'Azur, CNRS, Blvd de l'Observatoire, CS 34229, 06304 Nice cedex 4, France
  \label{OCA}
  \and
Laboratoire AIM, CEA/IRFU, CNRS/INSU, Universit\'e Paris Diderot, CEA-Saclay, 91191 Gif-Sur-Yvette, France 
  \label{CEA}
\and
Astronomy Instrumentation Group, University of Cardiff, UK
  \label{Cardiff}
\and
Institut d'Astrophysique Spatiale (IAS), CNRS and Universit\'e Paris Sud, Orsay, France
  \label{IAS}
\and
Institut N\'eel, CNRS and Universit\'e Grenoble Alpes, France
  \label{Neel}
\and
Institut de RadioAstronomie Millim\'etrique (IRAM), Granada, Spain
  \label{IRAME}
\and
Dipartimento di Fisica, Sapienza Universit\`a di Roma, Piazzale Aldo Moro 5, I-00185 Roma, Italy
  \label{Roma}
\and
Univ. Grenoble Alpes, CNRS, IPAG, F-38000 Grenoble, France 
  \label{IPAG}
    \and
Aix Marseille Universit\'e, CNRS, LAM (Laboratoire d'Astrophysique de Marseille) UMR 7326, 13388, Marseille, France
  \label{LAM}
\and
School of Earth and Space Exploration and Department of Physics, Arizona State University, Tempe, AZ 85287
  \label{Arizona}
\and
Universit\'e de Toulouse, UPS-OMP, Institut de Recherche en Astrophysique et Plan\'etologie (IRAP), Toulouse, France
  \label{IRAP}
\and
CNRS, IRAP, 9 Av. colonel Roche, BP 44346, F-31028 Toulouse cedex 4, France 
  \label{IRAP2}
\and
University College London, Department of Physics and Astronomy, Gower Street, London WC1E 6BT, UK
  \label{UCL}
\and 
Institut d'Astrophysique de Paris, Sorbonne Universit\'es, UPMC Univ. Paris 06, CNRS UMR 7095 75014, Paris, France
  \label{IAP}
\and 
LERMA, CNRS, Observatoire de Paris, 61 avenue de l'Observatoire, Paris, France
  \label{LERMA}
}

\date{Received \today \ / Accepted --}





\abstract
    {\emph{Context:} In the past decade, sensitive, resolved Sunyaev-Zel'dovich (SZ) studies of galaxy
      clusters have become common. Whereas many previous SZ studies have parameterized the pressure
      profiles of galaxy clusters, non-parametric reconstructions will provide insights
      into the thermodynamic state of the intracluster medium (ICM).
      
      \emph{Aims:} We seek to recover the non-parametric pressure profiles of the high redshift ($z=0.89$)
      galaxy cluster CLJ 1226.9+3332 as inferred from SZ data
      from the MUSTANG, NIKA, Bolocam, and Planck instruments, which all probe different angular scales.
     
      \emph{Methods:} Our non-parametric algorithm makes use of logarithmic interpolation,
      which under the assumption of ellipsoidal symmetry is analytically integrable.
      For MUSTANG, NIKA, and Bolocam we derive a non-parametric pressure profile 
      independently and find good agreement among the instruments. In particular, we find
      that the non-parametric profiles are consistent with a fitted gNFW profile.
      Given the ability of Planck to constrain the total signal,
      we include a prior on the integrated Compton Y parameter as determined by Planck.
      
      \emph{Results:}
      For a given instrument, constraints on the pressure profile diminish rapidly beyond the
      field of view. The overlap in spatial scales probed by these four datasets is
      therefore critical in checking for consistency between instruments.
      By using multiple instruments, our analysis of CLJ 1226.9+3332 
      covers a large radial range, from the central regions to the cluster outskirts:
      $0.05 R_{500} < r < 1.1 R_{500}$. This is a wider range of spatial scales
      than is typical recovered by SZ instruments.
      Similar analyses will be possible with the new generation
      of SZ instruments such as NIKA2 and MUSTANG2.}

\titlerunning{Non-parametric fitting with MUSTANG, NIKA, Bolocam, and Planck}
\authorrunning{C. Romero, M. McWilliam, and the NIKA SZ team}

\maketitle

\section{Introduction}
\label{sec:intro}


In recent years, Sunyaev Zel'dovich (SZ) effect observations have seen an increase in high resolution ($\theta \lesssim 30$\asecs)
observations \citep[e.g.][]{mason2010,adam2014,kitayama2016}. These observations come from MUSTANG on the
Robert C. Byrd Green Bank Telescope \citep[GBT][]{dicker2008}, NIKA on the IRAM 30-meter telescope \citep{monfardini2010},
and ALMA (band 3). However, all of these high resolution instruments have been limited in their ability to
recover signal beyond their field of view ($\sim 45$\asecs for MUSTANG and ALMA, and $\sim 120$\asecs for NIKA). As massive
galaxy clusters at moderate redshift ($z \sim 0.2-0.5$) have characteristic radii,
$R_{500} \gtrsim 3$\amin\footnote{$R_{500}$ is the radius within which the mean density is 500 times the
  critical density, $\rho_{cr}(z)$, of the universe, at the redshift, $z$, of the cluster.},
SZ observations made with these instruments have not been able to recover the
entire signal of the observed galaxy clusters. Therefore, observations from complementary SZ instruments which recover SZ at larger scales
such as Bolocam \citet{czakon2015} or Planck \citep{planck2013a} have been used in join analyses by
\citet{romero2015a} and \citet{adam2015,adam2016a} respectively.

These joint analyses have shown the ability to constrain the pressure profile of the intracluster medium (ICM) of individual
galaxy clusters over a large spatial range, often by assuming some parameterized pressure profile \citep[e.g.][]{romero2017,adam2014}.
In \citet{romero2015a}, the differences in fitted pressure profiles, especially in additional constraint of the inner pressure profile
slope, with the addition of MUSTANG data were noted. In the case of
\citet{romero2017,adam2015,adam2016a}, the pairs of instruments used did not have an overlap in recovered spatial scales, thus
limiting the ability to ascertain systematic errors of instruments.
However, as new SZ instruments like NIKA2 \citep[][]{monfardini2014,calvo2016} and MUSTANG2 \citep[][]{dicker2014a}
with the ability to recover a larger range of scales come online, there will be overlap. Consequently, for clusters observed
with multiple instruments (operating at different frequency bands), studies of the kinetic SZ effect, or relativistic
corrections \citep{itoh1998} will be of significant interest and stand to benefit from the additional frequency coverage, where
the additional frequency coverage will help with removal of contaminants (e.g. compact sources), as well as offering additional
leverage of the spectral distortion. 
To be sure of the results of these analyses, it will be critical to
understand any systematics involved with individual instruments. Recent results combining Bolocam and Planck data \citep{sayers2016},
which overlap in spatial scales recovered, show non-trivial changes (primarily of the outer slope of the pressure profile)
from previous Bolocam-only results \citep{sayers2013}.

Over a decade ago, the beta model
\citep{cavaliere1978} was favored; more recently other parameterizations such as a self-similar \citep{mroczkowski2009} and
analytic pressure profile based on a polytropic equation of state \citep{bulbul2010} have been explored. Of the parameterizations
of the ICM pressure profile, the generalized Navarro-Frenk-and-White \citep[gNFW][]{nagai2007} profile has garnered the most traction,
with a fairly canonical set of parameters coming from \citet{arnaud2010} (Hereafter, A10), which used a sample of 33 local ($z < 0.2$)
clusters.
Recently, several SZ studies have reconstructed non-parametric pressure profiles either through a maximum-likelihood approach
\citep[e.g.][]{ruppin2017,sayers2013} or deprojection of their data \citep[e.g.][]{basu2010,sayers2011}. The method employed in this
paper is a maximum likelihood approach, where the principle difference is our employment of analytic integrals.

Galaxy cluster formation is understood currently in the framework of hierarchical structure formation\citep[e.g.][]{press1974}.
While remarkable that a simple self-similar treatment of clusters \citep{kaiser1986} should describe the broad population of
galaxy clusters, non-linear physical processes in cluster formation (see \citet{kravtsov2012} for a review) likely account
for much of the scatter in scaling relations \citep[e.g.][]{battaglia2012}. In this context, investigating cluster pressure profiles
non parametrically can reveal deviations from a smooth pressure profile, which may correspond to departures from self-similarity
\citep{basu2010}. Moreover, these non-parametric fits do not rely on any physical model, and thus provide a less biased avenue
to constraining the thermodynamic state of the ICM. The combination of non-parametric SZ pressure profiles with complementary
non-parametric X-ray products, especially electron density, has \citep[e.g][]{basu2010,planck2013a,ruppin2017}
and will provide insights into the thermodynamic state of the ICM in clusters and likely be fundamental for improving cosmological
constraints via scaling relations. In fact, this is a significant motivation behind the NIKA2 tSZ large program \citep{comis2016},
as 300 hour program, using guaranteed time, to observe 50 homogeneously selected clusters at $z \gtrsim 0.5$.


Counts of galaxy clusters by mass and redshift serve to constrain cosmological parameters, notably the dark energy density
($\Omega_{\Lambda}$), matter density ($\Omega_{m}$), the amplitude of fluctuations ($\sigma_8$), and the equation of state of
dark energy ($w$) \citep{planck2016_szc}. Constraints on these
parameters derived from galaxy cluster samples are generally limited by the accuracy of mass estimation of 
galaxy clusters \citep[e.g.][]{hasselfield2013, reichardt2013}. Scaling relations which relate global (integrated) observables
to the cluster mass are often employed. Currently, scaling relations as applied to observables over an intermediate radial
region ($R_{2500} \lesssim r \lesssim R_{500}$) of galaxy clusters is preferred as this range shows minimal scatter in the
scaling relations \citep[e.g.][]{kravtsov2012}. This is due
to the generally low cluster-to-cluster scatter in pressure profiles, found observationally and in simulations,
within this radial range \citep[e.g.][]{borgani2004,nagai2007,arnaud2010,bonamente2012,planck2013a,sayers2013}.
While the relative homogeneity of pressure profiles in the intermediate region is well evidenced, it remains important to
develop methods to derive non-parametric pressure profiles of clusters so that physical deviations are not artificially
smoothed by the adoption of a parametric profile. 

The use of observables quantities determined at intermediate radii motivates the inclusion of instruments which are able
to recover the SZ signal out to these radii, while the need to recover deviations from a smooth profile favor higher
resolution instruments. In order to then cover a wide range of angular scales ($0.05 R_{500} < r < 1.1 R_{500}$), we have
performed fits on MUSTANG, NIKA and Bolocam data, with the addition of a prior from Planck data.
This paper is organized as follows. In Section~\ref{sec:obs} we review the NIKA, MUSTANG, and Bolocam observations and reduction. 
In Section~\ref{sec:ml_deproj} we address the method used to non-parametrically fit pressure profiles to each of the data sets.
We present results from our non-parametric fits in Section~\ref{sec:np_res} and parametric fits in Section~\ref{sec:parfits}. 
Throughout this paper we assume a $\Lambda$CDM cosmology with $\Omega_m = 0.31$, $\Omega_{\lambda} = 0.69$, and $H_0 = 68$ 
km s$^{-1}$ Mpc$^{-1}$, consistent with the cosmological parameters derived from the full \emph{Planck} mission
\citep{planck2016_cp}. For this cosmology, at $z=0.89$, we have a scale of 7.945 kpc/\asec.

\section{Observations and Data Reduction}
\label{sec:obs}

\subsection{CLJ1226.9+3332}
\label{sec:sample_clj1227}



At a high redshift of $z=0.89$, CLJ1226.9+3332, hereafter CLJ 1227, is a massive cluster which was first discovered in the
Wide Angle ROSAT Pointed Survey \citep[WARPS][]{ebeling2001}. It has successively been well studied in the X-ray
\citep[\emph{XMM},\emph{Chandra}, and \emph{XMM/Chandra}][respectively]{maughan2004,bonamente2006,maughan2007}
and SZ \citep[][]{joy2001,muchovej2007,mroczkowski2009,mroczkowski2011,bulbul2010,korngut2011,adam2015}.
In \citet{maughan2007}, the identification of hot southwestern component gave the first indications of disturbance
in this cluster. This interpretation was further bolstered by HST observations \citep{jee2009a}, in which the lensing
analysis revealed two distinct peaks, one of which was coincident with the hot X-ray temperature region.

From the first SZ measurements of CLJ 1227 with BIMA \citep[][]{joy2001} has generally appeared
azimuthally symmetric and relaxed. Later studies with SZA \citep{muchovej2007,mroczkowski2009,mroczkowski2011}
all appear to re-affirm this symmetry, while the evidence in SZ observations for a potential disturbance
in the core region begins to grow. \citet{korngut2011} find a ridge of significant substructure in
MUSTANG data, which when compared with X-ray profiles, is consistent with a merger scenario within
CLJ 1227. However in the current processing of MUSTANG data \citep{romero2017},
this substructure is not evident. Combining the SZ pressure profile with X-ray electron density profile,
\citet{adam2015} find relatively large entropy values in the core as support for disturbance on small scales.
A similar conclusion is reached by \citet{rumsey2016}, who find that the core
of CLJ 1227 exhibits signs of merger activity, while the outskirts appear relaxed.

Given the relative circular symmetry of CLJ 1227, it provides a suitable test for determining a non-parametric pressure
profile of the cluster, while maintaining the assumption of spherical symmetry. For the centroid, we adopt the X-ray
centroid from ACCEPT \citep{cavagnolo2009} is at [RA,Dec] = [12:26:57.9,+33:32:49] (J2000).
From X-ray data, \citet{mantz2010b} determined a scale radius $R_{500} = 1000 \pm 50$ kpc, which corresponds to
$M_{500} = (7.8 \pm 1.1) \times 10^{14} M_{\odot}$. In the following, we summarize how the data, which are used in this study,
were produced in previous studies. In Table~\ref{tbl:cluster_obs}, we summarize the angular scales probed by the instruments
and the overall depth of observations.

\begin{figure*}[!h]
  \centering
  \includegraphics[width=0.33\textwidth]{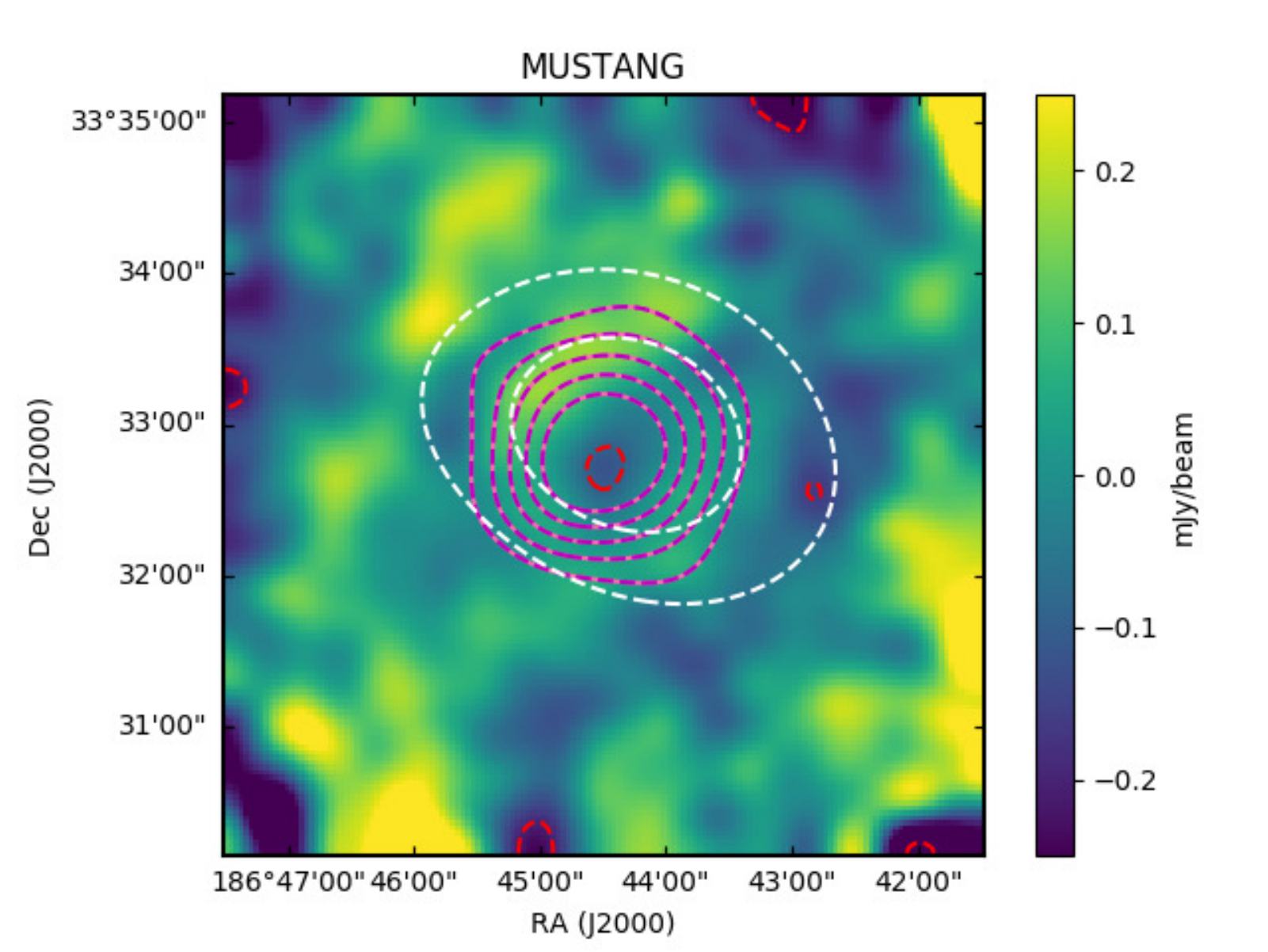}
  \includegraphics[width=0.33\textwidth]{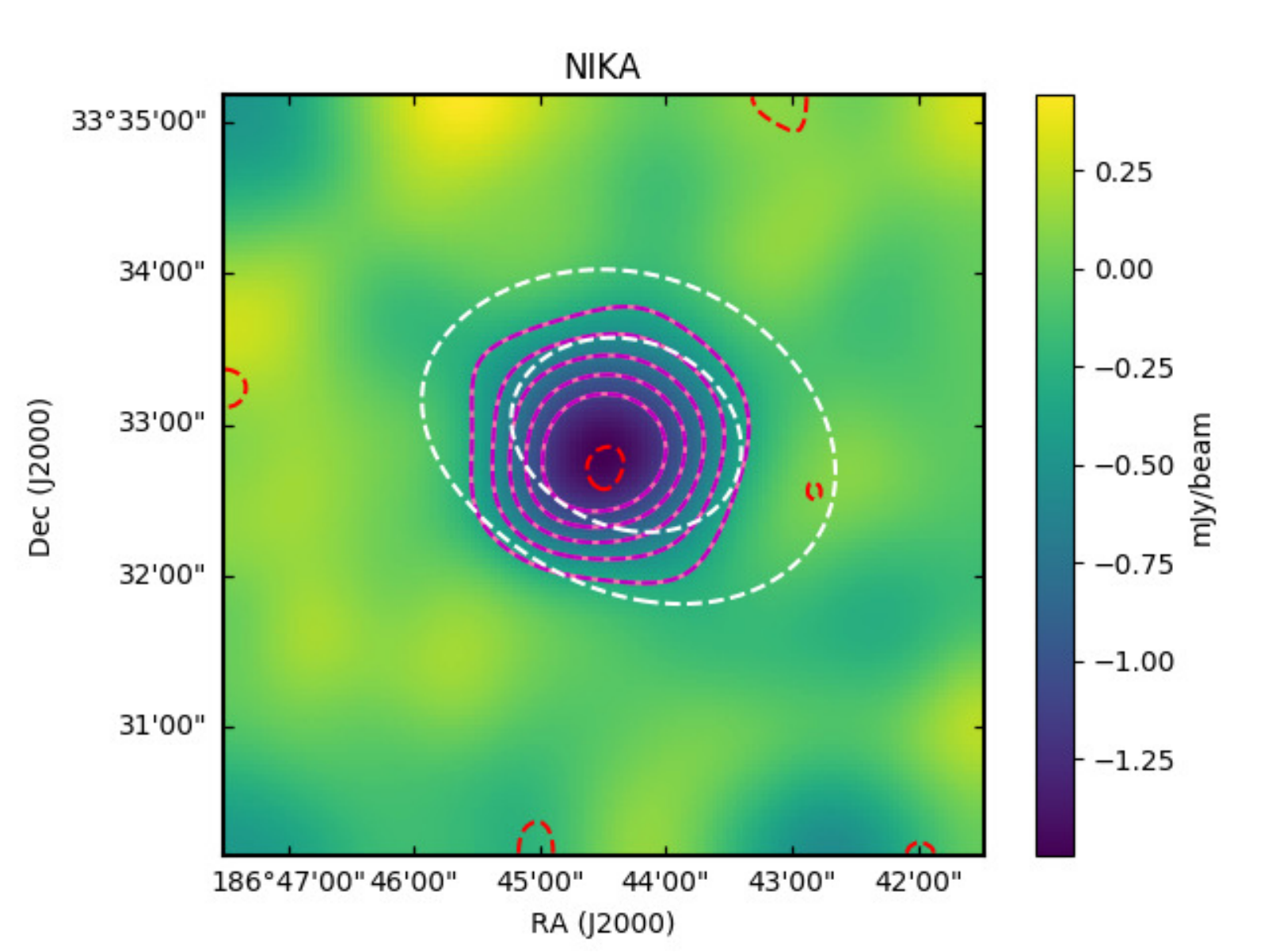}
  \includegraphics[width=0.33\textwidth]{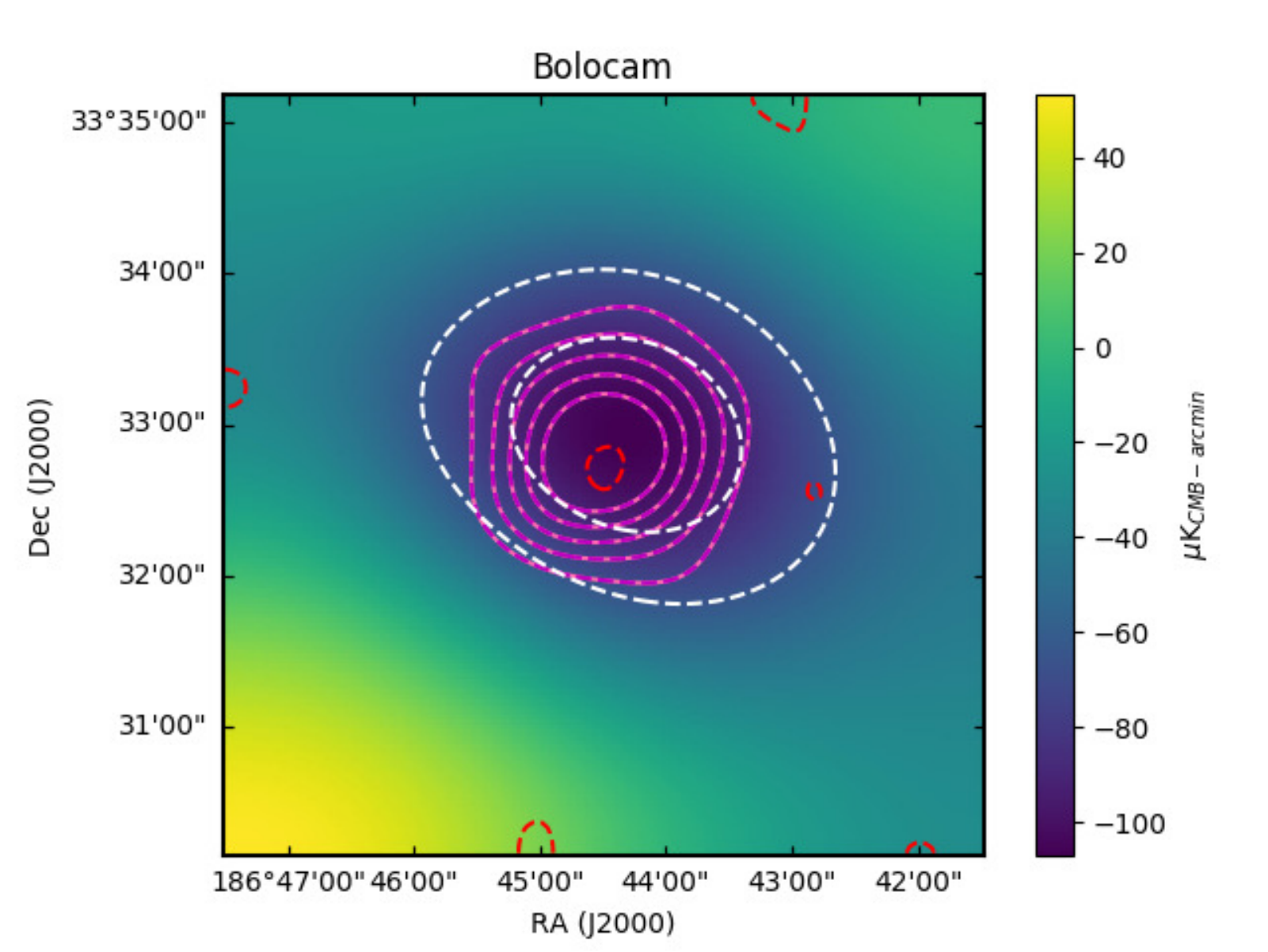}
  \caption{Left: MUSTANG map, smoothed by a 10\asecs FWHM kernel; Middle: NIKA (2mm) map smoothed by a 10\asecs FWHM kernel;
    Right: Bolocam map smoothed by a 60\asecs FWHM kernel. In all three panels, the red contours are those
    of MUSTANG, magenta contours of those of NIKA, and white contours are those of Bolocam. For MUSTANG and Bolocam, the contours
    start at $(-)2\sigma$, with $1\sigma$ increments. For NIKA, the contours start at $(-)3\sigma$ with $2\sigma$ increments.
    The point source identified in \citet{adam2015} is subtracted in the MUSTANG
    and NIKA maps.}
  \label{fig:clj1227_maps}
\end{figure*}





\begin{table*}[]
  \caption{\footnotesize{Overview of instrument parameters influencing the constraining power of
      pressure profiles for CLJ 1227 relevant for this analysis. Noise is determined on maps smoothed
      by a Gaussian kernel with FWHM of 10\asec, 10\asec, and 60\asec for MUSTANG, NIKA, and Bolocam
      respectively.}}
\begin{center}
\begin{tabular}{l|lllll}
  \hline
  \hline \\
  Instrument & Freq. & $T_{obs}$ & Noise & FWHM & FOV \\
   & (GHz) & (hours) & (Compton y) & (\asec) & (\amin) \\
  \hline \\
  \textbf{MUSTANG} &  90 & 4.9  & $34.2 \times 10^{-6}$ & 9   & 0.7 \\
  \textbf{NIKA}    & 150 & 7.2  & $12.5 \times 10^{-6}$ & 18  & 1.9 \\
  \textbf{Bolocam} & 140 & 11.8 & $8.48 \times 10^{-6}$ & 58  &   8 \\
  \hline
\end{tabular}
\end{center}
\label{tbl:cluster_obs}
\end{table*}

\subsection{Overview of MUSTANG data products}
\label{sec:musobs}
The MUSTANG camera \citep{dicker2008}, while on the 100 meter Robert C. Byrd Green Bank Telescope
\citep[GBT, ][]{jewell2004}, had angular resolution of 9\asec (full-width, half-maximum FWHM) and was one of only
a few SZ effect instruments with sub-arcminute resolution. With a pixel spacing of 0.63$f \lambda$, MUSTANG's
instantaneous field of view (FOV) is 42\asecs, and is limited in its ability to recover scales larger than $\sim45$\asec.
MUSTANG is a 64 pixel array of Transition Edge Sensor (TES) bolometers arranged in an $8 \times 8$ array
and had been located at the Gregorian focus of the 100 m GBT. Operating at 90 GHz (81--99~GHz),
The conversion factor from Jy/beam to Compton parameter used is $\mhyphen 2.50$, including relativistic
corrections using an isothermal electron temperature from X-ray data $k_B T_x = 12$ keV \citep{sayers2013}.
The calibration and pointing uncertainties are 10\% and 2\asec respectively \citet{romero2017}.
More detailed information about the instrument can be found in \citet{dicker2008}. 

The observations and data reduction are described in detail in \citet{romero2015a}, and were applied to the MUSTANG
data of CLJ 1227 as presented in \citet{romero2017}.
The MUSTANG data map, with a point source subtracted (see Section~\ref{sec:preprocessing})
is shown in the left panel in Figure~\ref{fig:clj1227_maps}.

For this analysis, we refine the transfer function found in  \citet{romero2017} by filtering a cluster model
using a strictly \citetalias{arnaud2010} profile (a gNFW profile with parameters
$[\alpha,\beta,\gamma,C_{500},P_0]=[1.05,5.49,0.31,1.18,8.42 P_{500}]$) through the standard MUSTANG pipeline.
The resultant transfer function is then merged with the prior transfer function \citep[on white noise][]{romero2017}.
The principle difference between this new transfer function and the former one
occurs at scales larger than the FOV (angular frequencies smaller than $\sim0.025$ inverse arcseconds).
We check the robustness of the transfer functions to the standard pipeline across a range of cluster models
(gNFW profiles with varying parameters) and find agreement, principally of the peak amplitude, within 10\%.

Moreover, we verify the fidelity of the new transfer function by reproducing the analysis performed in
\citet{romero2017} for CLJ1227, with the use of the new transfer function in place of the standard MUSTANG
filtering procedure. We find good agreement with the previous results, where
f
the best fit profile shape parameters ($C_{500}$, $P_0$, and $\gamma$ - see Section~\ref{sec:parfits})
are within $\sim10$\% agreement of the values reported in \citet{romero2017}.

\subsection{Overview of NIKA data products}
\label{sec:nikaobs}

NIKA \citep{monfardini2010,monfardini2014} was a dual band camera working at 150 and
260 GHz, and consisted of 253 Kinetic Inductance Detectors (KIDs) operating at 100 mK by using a closed cycle $^3$He-$^4$He dilution fridge. 
Furthermore, with a sensitivity of 14 (35) mJy/beam.s$^{1/2}$ , a circular field-of-view (FOV) of 1.9$^{\prime}$ (1.8$^{\prime}$), and a
resolution of 18.2$^{\prime \prime}$ (12.0$^{\prime \prime}$) at 150 (260) GHz NIKA was particularly well adapted to map the thermal
Sunyaev-Zeldovich effect in such a high redshift cluster. Including calibration (7\% and 12\%) and bandpass uncertainties, the NIKA conversion
factors from Jy/beam to Compton parameter are $-10.9 \pm 0.8$ and $3.5\pm0.5$ at 150 and 260 GHz, respectively. The pointing RMS
achieved during CLJ 1227 observations was below 3\asec.
A detailed description of the general performances of the camera can be found in \citet{catalano2014,adam2014}.

In this analysis, we employ NIKA camera data of the cluster CLJ 1227, which were obtained at the IRAM 30 m telescope (Pico Veleta)
in February 2014, processed with the NIKA processing pipeline described in \citet{adam2014}, and  presented in \citet{adam2015}.
CLJ 1227 was mapped using on-the-fly raster scans with an on-cluster time of 7.8 hours.
The transfer function of the processing procedure, which is used in this analysis, was computed using signal plus noise simulations as
described in \citet{adam2015}.
Overall the transfer function is consistent with a constant value of 0.95 for angular scales smaller than the NIKA FOV and larger
than the size of the NIKA beam. Using the 260~GHz NIKA map, \citet{adam2015} identified a point source located 30$^{\prime \prime}$
southeast of the center of the cluster. The 150~GHz NIKA map, with the point source subtracted (Section~\ref{sec:preprocessing})
is shown in the middle panel in Figure~\ref{fig:clj1227_maps}.

\subsection{Overview of Bolocam data products}
\label{sec:bolocamobs}

To probe a wider range of scales we complement the MUSTANG and NIKA data with SZ data from Bolocam \citep{glenn1998}. 
Bolocam is a 144-element bolometer
array on the Caltech Submillimeter Observatory (CSO) with a beam FWHM of 58\asecs at 140 GHz and circular FOV with 8\amins 
diameter \citep{glenn1998,haig2004}, which is well matched to the angular size of $R_{500}$ ($\sim 2$\amin) of CLJ 1227. 
Bolocam's conversion factor to Compton y from $\mu$K$_{CMB}$ is reported as $-3.69 \times 10^{-7}$, with the relativistic
corrections ($k_B T = 12$ keV) taken into account.

Bolocam was a facility instrument on the CSO from
2003 until 2012. CLJ 1227 was observed with a Lissajous pattern that results in a tapered
coverage dropping to 50\% of the peak value at a radius of roughly 5\amin, and to 0 at a radius of 10\amin.
The Bolocam maps used in this analysis are $14\arcmin \times 14\arcmin$. The Bolocam data 
are the same as those used in \citet{czakon2015} and \citet{sayers2013}; the details of the reduction are 
given therein, along with \citet{sayers2011}. The Bolocam map is shown in the right panel of Figure~\ref{fig:clj1227_maps}.
The reduction and calibration is similar to that used for MUSTANG, and Bolocam achieves a 
5\% calibration accuracy and 5\asecs pointing accuracy.


\subsection{Planck integrated Compton parameter}
\label{sec:picp}

As in \citet{adam2015}, we wish to include constraints on still larger scales than reached with the aforementioned instruments,
and therefore we include in this analysis the integrated Compton parameter of the cluster as measured using Planck data.
We use the Planck frequency maps from 143 to 857~GHz to produce a Compton parameter map as described in \citet{hurier2013} and
\citet{planck2013ymap,planck2016_tsz}. The resolution of this map is 7.5$^{\prime}$, limited by the lowest frequency Planck channel
map used in the reconstruction. Using this map we compute the integrated Compton parameter up to a radial distance of 15$^{\prime}$
Uncertainties in the integrated Compton parameter are computed by integrating at random positions
around the cluster. The uncertainties obtained have been also crossed-checked using Planck half-ring half difference Compton parameter
map
obtained as described in \citet{planck2013ymap,planck2016_tsz}. We find $Y_{Cyl,}(15^{\prime}) = (0.94 \pm 0.36) \times 10^{-3}$ arcmin$^2$. 


\section{Non-parametric Pressure Profile Reconstruction via a Maximum Likelihood Analysis}
\label{sec:ml_deproj}


We perform non-parametric fits of the pressure profile of CLJ 1227 on MUSTANG, NIKA, and Bolocam data maps independently.
The filtering effects incurred from data processing of these instruments favors forward modelling pressure profiles and
performing fits over trying to do a geometric deprojection, for example with Abel transforms \citep[e.g.][]{basu2010}.
Before fitting the pressure profiles (Section~\ref{sec:preprocessing}),
we remove a point source from the MUSTANG and NIKA maps based on previous works
\citep{adam2015,romero2017}. Additionally, we ensure that a mean level has been removed in the MUSTANG and BOLOCAM maps. 
The construction of our non-parametric galaxy cluster model is described in Section~\ref{sec:np_models}, the fitting
procedure is described in Section~\ref{sec:fitting}, and we review the performed validity checks in Section~\ref{sec:validity}.

\subsection{Preprocessing}
\label{sec:preprocessing}


A point source at $4.6\sigma$ significance ($\sim 0.5$ mJy) in MUSTANG was reported in \citet{korngut2011}, but is not
evident in the MUSTANG data as reprocessed in \citet{romero2017}. A short VLA filler observation (VLA-12A-340, D-array, at 7 GHz)
was performed to follow up this potential source (at RA 12:26:58.0 and Dec +33:32:59), but to a limit of $\sim 50 {\rm \mu Jy}$ nothing
is seen\citep{romero2017}. At 500 $\mu$m, \emph{Herschel}-SPIRE has a point source sensitivity of $\sim 8$ mJy,
and therefore does not truly constrain a potential point source at this location.

\citet{adam2015} find a point source at a different location, RA 12:26:59.855 and Dec +33:32:35.21, with a flux density of 
$6.8 \pm 0.7 \text{ (stat.)} \pm 1.0 \text{ (cal.)}$ mJy at 260 GHz and $1.9 \pm 0.2 \text{ (stat.)}$ at 150 GHz.
For this source, at 500 $\mu$m, \emph{Herschel}-SPIRE finds a flux of $100 \pm 8$ mJy.
\footnote{{http://irsa.ipac.caltech.edu/applications/Gator/}} A point source at this location is fit to the MUSTANG
data with a flux density of $0.36 \pm 0.11$ mJy \citep{romero2017}.
We subtract this point source from the NIKA and MUSTANG maps using the above flux density values.
In the Bolocam data, the point source is faint enough to not be a concern, given Bolocam's beam size.


We also wish to account for any mean level before fitting our cluster model, especially because there is a degeneracy
between the mean level and the cluster model, and the mean level can typically be well constrained a priori.
The mean level in the MUSTANG map is calculated
as the mean within the inner arcminute MUSTANG noise map, which was created from time-flipped time-ordered data.
We subtract the mean level from the MUSTANG map before fitting a cluster model.
Within NIKA data, a mean level is calculated within the timestreams for data falling outside
the masked region. This mean level is subtracted within the timeline processing of NIKA data. The Bolocam map already has
a mean level subtracted.

\subsection{Non-Parametric Pressure Profile Models}
\label{sec:np_models}

Our non-parametric pressure profile reconstruction assumes spherical symmetry and power law interpolation between radial bins.
Because we employ analytic integrals, we can integrate from zero to a finite radius, and from a finite radius to infinity,
with some clear restrictions on the power
laws when doing so. The analytic integration has been employed before \citep[e.g.][]{vikhlinin2001a,korngut2011,sarazin2016}.
Here, we resolve  previous limitations (Appendix~\ref{sec:analytic_integrals}) found with certain power laws for which the
previously given analytic formulation are undefined \citep{korngut2011,sarazin2016}, but which are necessary to be covered
in our analysis.
Our fitting algorithm is applied
to each dataset independently; therefore, cluster models are binned and gridded differently for each dataset.
Radial bins are defined so that
each bin is at least as wide as a beam width (FWHM), with the additional constraint that the outer most bin
is beyond the FOV of the instrument.

For each bin, $i$, we denote the radius as $R_i$, and assign a pressure $P_i$. The interpolation of pressure
between at a radius $r$, $R_i < r < R_{i+1}$ is given by $P(r) = P_i (r/R_{i})^{-\alpha}$, where $\alpha$ is
calculated as:
\begin{equation}
  \alpha = -\frac{\log(P_{i+1}) - \log(P_i)}{\log(R_{i+1}) - \log(R_i)}.
\end{equation}
For radii interior to our innermost radial bin ($R_1$), we extrapolate using the same power law as between $R_1$ and
$R_2$. Similarly, for radii exterior to our outermost radial bin ($R_n$), we extrapolate using the same power law as
between $R_{n-1}$ and $R_n$. We therefore put a prior on our outermost slope such that $\alpha > 1$, and the integrated
quantity is finite.

We note that for a non-rotating, spherical object in hydrostatic equilibrium (HSE) under entirely thermal pressure
support, the power law should be
limited to $\alpha >4$ in order to avoid having infinite mass (see Appendix~\ref{sec:finite_mass}).


Given the restrictions of ellipsoidal symmetry and a power law dependence of the integrated quantity (pressure) on the
ellipsoidal radius, it is possible to calculate the integral along the line of sight analytically
\citep[e.g.][]{vikhlinin2001a,korngut2011}. We follow
principally the formulation provided in \citet{korngut2011}. As noted in \citet{sarazin2016}, there are certain power
laws ($\alpha/2 = p = 1/2, 0 , -1/2, -1, -3/2, ...$) for which the formulation given in \citet{korngut2011} fails,
but a reformulation provides a valid integration. More generally, the formulation fails for $\alpha/2 = p < -1/2$.
While a negative index indicates a rise in pressure with radius (atypical), this could arise, especially localised,
from shocks, for example. We also wish to minimise our restrictions on the power laws (between bins) so as to minimise
induced correlations between bins. Therefore, we implement extensions to the canonical formulation that allow us to
integrate within finite regions (spheres or shells that extend only to a finite radius). These extensions and
reformulations of specific half integers are described in Appendix~\ref{sec:analytic_integrals}.

The integrated profiles, calculated as the Compton y parameter:
\begin{equation}
  y = \frac{\sigma_T}{m_e c^2} \int P_e dl,
  \label{eqn:compton_y}
\end{equation}
are converted into the units of the original data map. Maps are gridded by assuming a linear interpolation
of the 1D (radial) profiles. When gridding our bulk ICM component, we adopt the ACCEPT centroid of CLJ 1227,
and grid a larger map than used for fitting.
These maps are then convolved with the respective beam and transfer function. Aliasing is aleviated by trimming
the region not fitted. For MUSTANG, NIKA, and Bolocam, we fit a square region about the centroid with lengths
of 2\amin, 4\amin, and 13.33\amin respectively.


\subsection{Fitting Algorithm}
\label{sec:fitting}

We employ a maximum likelihood algorithm, and take our noise to be Gaussian.
In previous works, NIKA and MUSTANG noise have been taken as uncorrelated \citep[e.g.][]{romero2015a,romero2017,adam2015}.
Bolocam noise has been taken as approximately uncorrelated, but 1000 noise realizations, which included CMB and point
source estimates, are provided to allow for a more accurate noise estimation \citep{sayers2011}.
We calculate the two-dimensional power spectrum for noise maps of each dataset and find that the noise is consistent
with white noise on the scales, for each dataset, which we wish to constrain the models.

We calculate the final probability of our models by applying priors as prescribed by Bayes' Theorem.
On each of the pressure bins, we assign strict priors that $P_i > 0$, and as previously mentioned, the last bin
has a prior that on its associated power law slope: $\alpha > 1$. We allow for the choice of including a prior
on the integrated Compton Y parameter:
\begin{equation}
  Y = \int y d\Omega,
  \label{eqn:integrated_y}
\end{equation}
where the integral over solid angle taken within a given radius is generally referred to as the cylindrical
Compton Y value ($Y_{cyl}$). We calculate $Y_{cyl}$ using the un-filtered Compton y profile (before convolution with
an instrument's beam and transfer function). The prior on Y comes
from Planck data \citep{planck2014} as discussed in Section~\ref{sec:picp}.
In particular, we take the prior $Y_{cyl}(15$\amin$) = (0.94 \pm 0.36) \times 10^{-3} \text{arcmin}^2$.

We employ the described probability function in a python Markov Chain Monte Carlo (MCMC) package, emcee \citep{foreman2013}.
This MCMC package makes use of a variant of a Metropolis-Hastings (MH) algorithm, in particular the ensemble sampling algorithm is
affine-invariant \citep{goodman2010}. The use of ensemble sampling, as opposed to a canonical single-point sampling, contributes
to the notable advantage of this algorithm (within emcee) having a much shorter autocorrelation time than a standard MH algorithm.
Furthermore, the computationally expensive part of drawing a new walker has been parallelized.


\subsection{Validation and Performance of Fitting Algorithm}
\label{sec:validity}

Our algorithm is first tested with mock cluster observations. We create mock observations by adding
a noise realization (created from jack-knifed timestreams) for each of the three maps to the
corresponding filtered map of a previously determined \citep{romero2017} gNFW profile.
We perform initial tests to validate the number of bins chosen. The resolution (FWHM) and FOV of our
instruments (see Section~\ref{sec:np_models}) suggest that between 4 and 8 bins are appropriate.
We cover this range, with fits run with 4, 6, and 8 bins with two model constructions. The first,
as described in Section~\ref{sec:np_models}, and a second being uniform spherical pressure bins.
We quantify the performance in terms of the logarithmic likelihoods ($\ln (\mathcal{L}) = -\chi^2 / 2$)
of (1) the fit to fitted model maps to the mock observations, and (2) the fits of the non-parametric
fits to the input pressure profile. In combination, we find that $\ln (\mathcal{L})$ typically varies
by $<2$, and that 6 bins with the power law model perform best for MUSTANG and Bolocam, while 4 bins
with the power law performs best for NIKA. We therefore adopt the 6-bin power law approach as our
standard method of modeling the Compton y map.

We further test the dependence of the fit results on initial guess of the pressure values, and find that this
dependence is minimal. We change the input guesses by the
following factors $f_P = [0.01,0.1,0.33,3.0,10,100]$, and perform the fits on the mock cluster observations.
We find that at worst, we see that the results are generally within 7\% of each other, with the exception
that the outermost bin may see a dispersion up to 20\%, and one of the inner bins in NIKA data sees a disperion
of 14\%. However, if we limit the span to just $f_P = [0.1,0.33,3.0,10]$, the dispersions are less than 6\%
for all but the outer bins, which see dispersions less than 10\%.

Finally, across the above suite of tests (number of bins, uniform or power law distribution within a bin, and
initial guesses), we find that the outermost bins in MUSTANG and Bolocam are biased high, where for Bolocam,
the second most outer bin is also biased high. We find that in the production of the models, this appears to arise
with the application of the transfer functions of these instruments. As we define the outermost bin of our power
law model to extend to infinity, truncating, or reducing the number of bins does not resolve this bias. Rather,
we find it best to retain the bins in the map fitting procedure and to trim them in subsequent analyses. Therefore,
in our analysis of real data, 6 bins are used within the fitting procedure, and we retain 5, 6, and 4 bins for
MUSTANG, NIKA, and Bolocam respectively for subsequent analysis and discussion.

\begin{figure}[!h]
  \centering
  \includegraphics[width=0.5\textwidth]{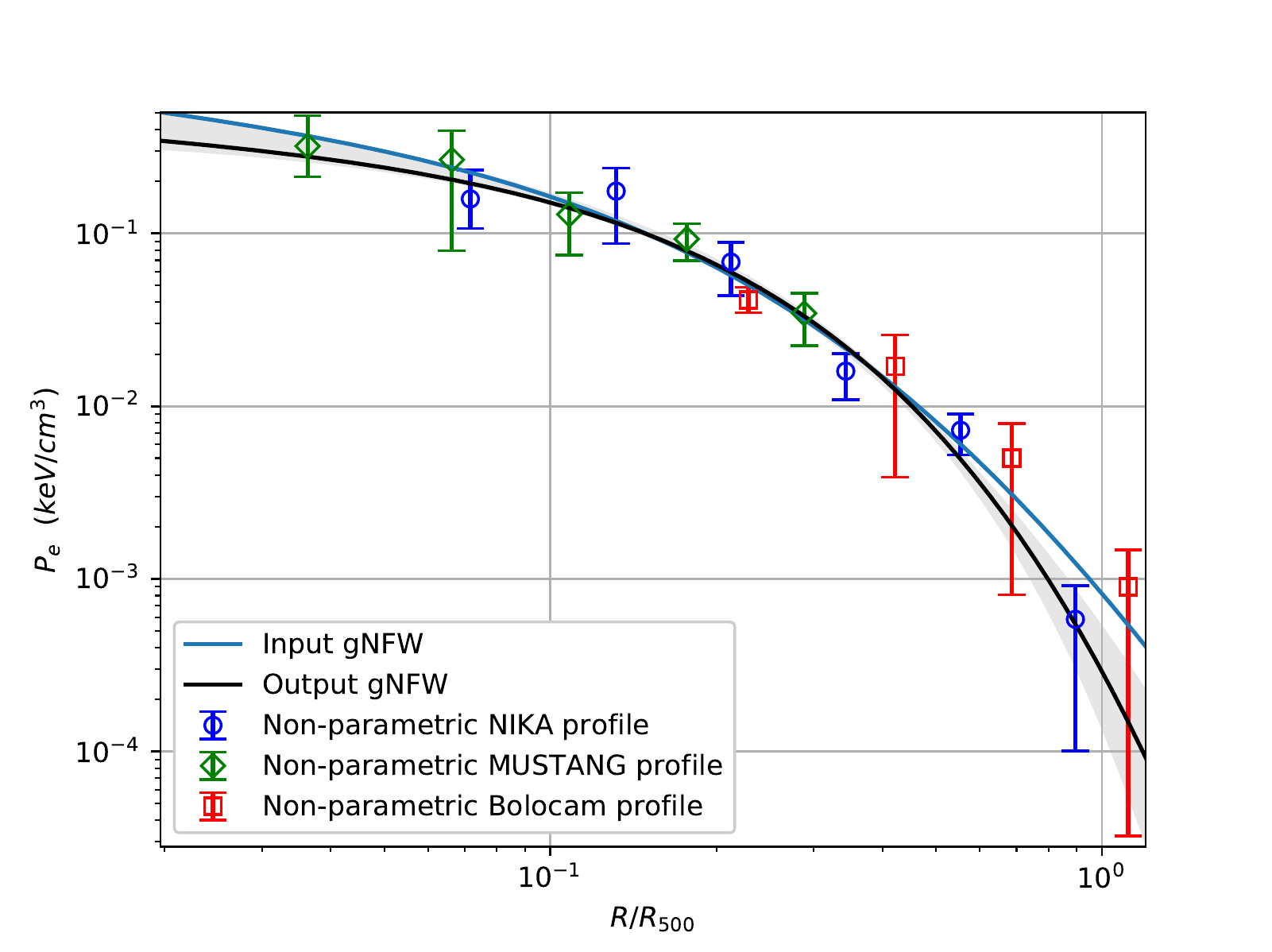}
  \caption{Non-parametric pressure profiles as determined via each mock observation individually,
    and the gNFW (parametric) pressure profile as simultaneously fit to the non-parametric pressure profiles.
    The error bars are statistical, from the MCMC fits.}
  \label{fig:virt_robustness}
\end{figure}

We show the fits to our mock observations with this 6-bin, power-law model in Figure~\ref{fig:virt_robustness},
and note that the reconstructed profiles are consistent with the input profile. Our input gNFW model has been taken
from \citet{romero2017}, and our output model is fit as prescribed in Section~\ref{sec:parfits}.




\section{Non-Parametric Pressure Profile Results}
\label{sec:np_res}

\begin{figure}[!h]
  \centering
  \includegraphics[width=0.5\textwidth]{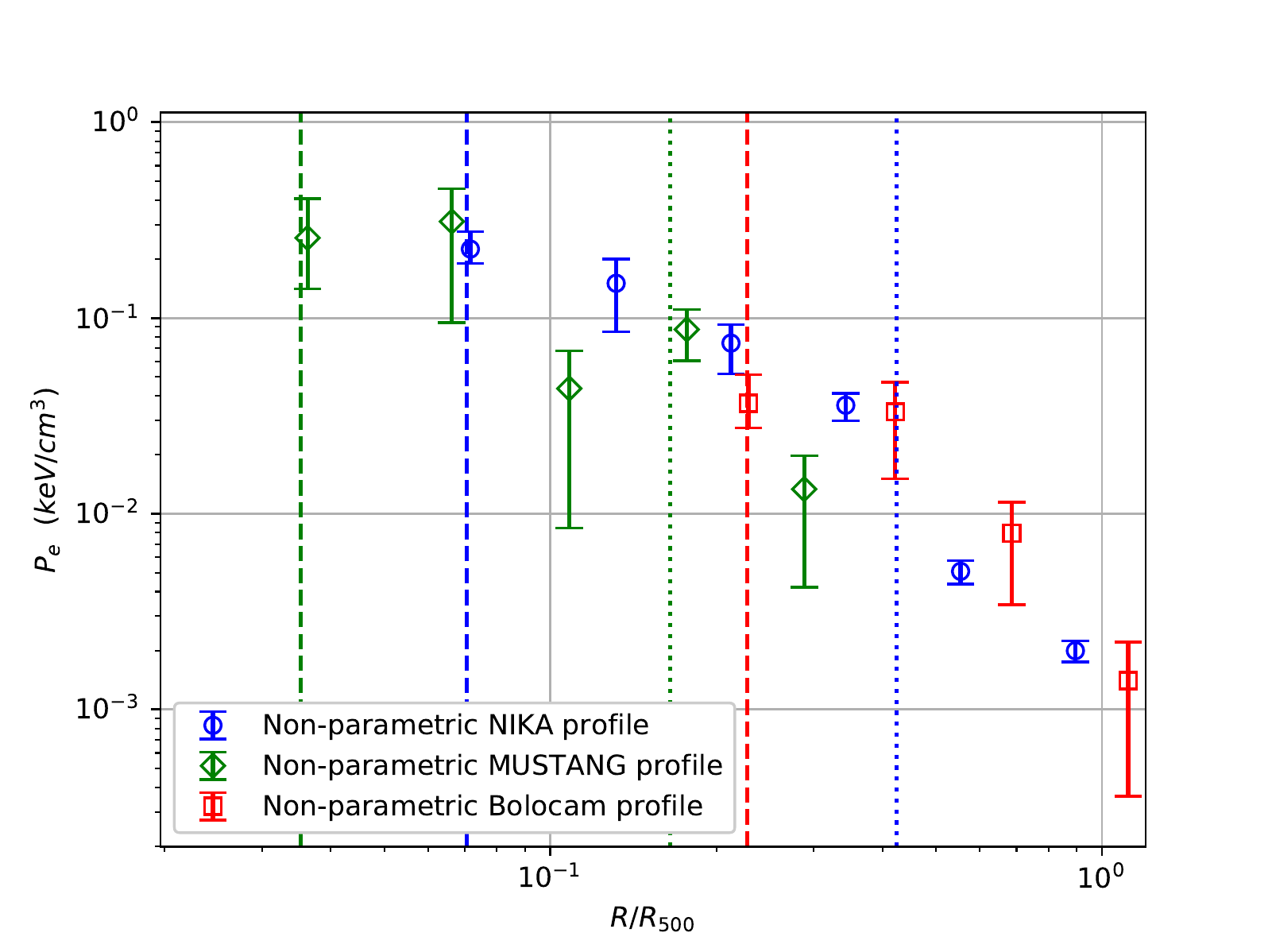}
  \caption{Non-parametric pressure profiles as determined via each dataset individually. The error bars are statistical,
    from the MCMC fits. The vertical dashed and dotted lines correspond to the half width at half maximum (HWHM) and
    FOV/2 (i.e. radial FOV), respectively.}
  \label{fig:nppps}
\end{figure}

Given the MUSTANG transfer function, we expect that the constraints beyond 42\asecs (radially) are negligible. Therefore,
we exclude the outermost radial bin from further analysis.
The Bolocam transfer function is provided as a two-dimensional transfer function. We find that the transfer function produces
artifacts at large radii ($r \gtrsim 1000$ kpc) for all plausible cluster models. While we find it important to include these
outer bins for the fitting procedure itself, we exclude the outer two bins from further analysis. The results, after these
exclusions, are tabulated in Figure~\ref{fig:nppps} and shown in Table~\ref{tbl:nppp_res}.
\begin{table}  
  \caption{\footnotesize{Non parametric pressure profile fits.} The red rows
    correspond to bins which have been trimmed.}
  \begin{center}
    \begin{tabular}{|l|lll|}
      \hline
      R     & $P_e$          & $\sigma_{P_e,low}$ & $\sigma_{P_e,high}$ \\
      (kpc) & (keV cm$^{-3}$) & (keV cm$^{-3}$)   & (keV cm$^{-3}$)   \\
      \hline
      \multicolumn{4}{|c|}{NIKA} \\
      \hline
      73    & 0.225   & 0.051   & 0.035   \\
      134   & 0.150   & 0.049   & 0.065   \\
      216   & 0.0744  & 0.0181  & 0.0226  \\
      349   & 0.0358  & 0.0053  & 0.0060  \\
      564   & 0.00508 & 0.00068 & 0.00071 \\
      910   & 0.00200 & 0.00024 & 0.00024 \\
      \hline
      \multicolumn{4}{|c|}{MUSTANG} \\
      \hline
      37   & 0.257    & 0.151   & 0.115   \\
      67   & 0.311    & 0.146   & 0.216   \\
      110  & 0.0436   & 0.0243  & 0.0352  \\
      180  & 0.0874   & 0.0231  & 0.0270  \\
      294  & 0.0133   & 0.0064  & 0.0091  \\
      \textcolor{red}{479}  & \textcolor{red}{0.000959} &
      \textcolor{red}{0.00082} & \textcolor{red}{0.00284} \\
      \hline
      \multicolumn{4}{|c|}{Bolocam} \\
      \hline
      233  & 0.0367   & 0.0147   & 0.0092  \\
      429  & 0.0332   & 0.0139   & 0.0181  \\
      698  & 0.00795  & 0.0035   & 0.0045  \\
      1135 & 0.00141  & 0.00081  & 0.00104 \\
      \textcolor{red}{1845} & \textcolor{red}{0.00320}  &
      \textcolor{red}{0.00083}  & \textcolor{red}{0.00084} \\
      \textcolor{red}{3000} & \textcolor{red}{0.00101}  &
      \textcolor{red}{0.00044}  & \textcolor{red}{0.00047} \\
      \hline
    \end{tabular}
  \end{center}
  \label{tbl:nppp_res}
\end{table}



From the Monte Carlo chains of the non-parametric fits, we determine the covariance matrix of the pressure bins for each dataset as:
\begin{equation}
  \mathbf{N}_{i,j} = \langle d_i d_j \rangle - \langle d_i \rangle \langle d_j \rangle.
  \label{eqn:covariance}
\end{equation}

\begin{figure*}[h]
  \centering
  \includegraphics[width=0.33\textwidth]{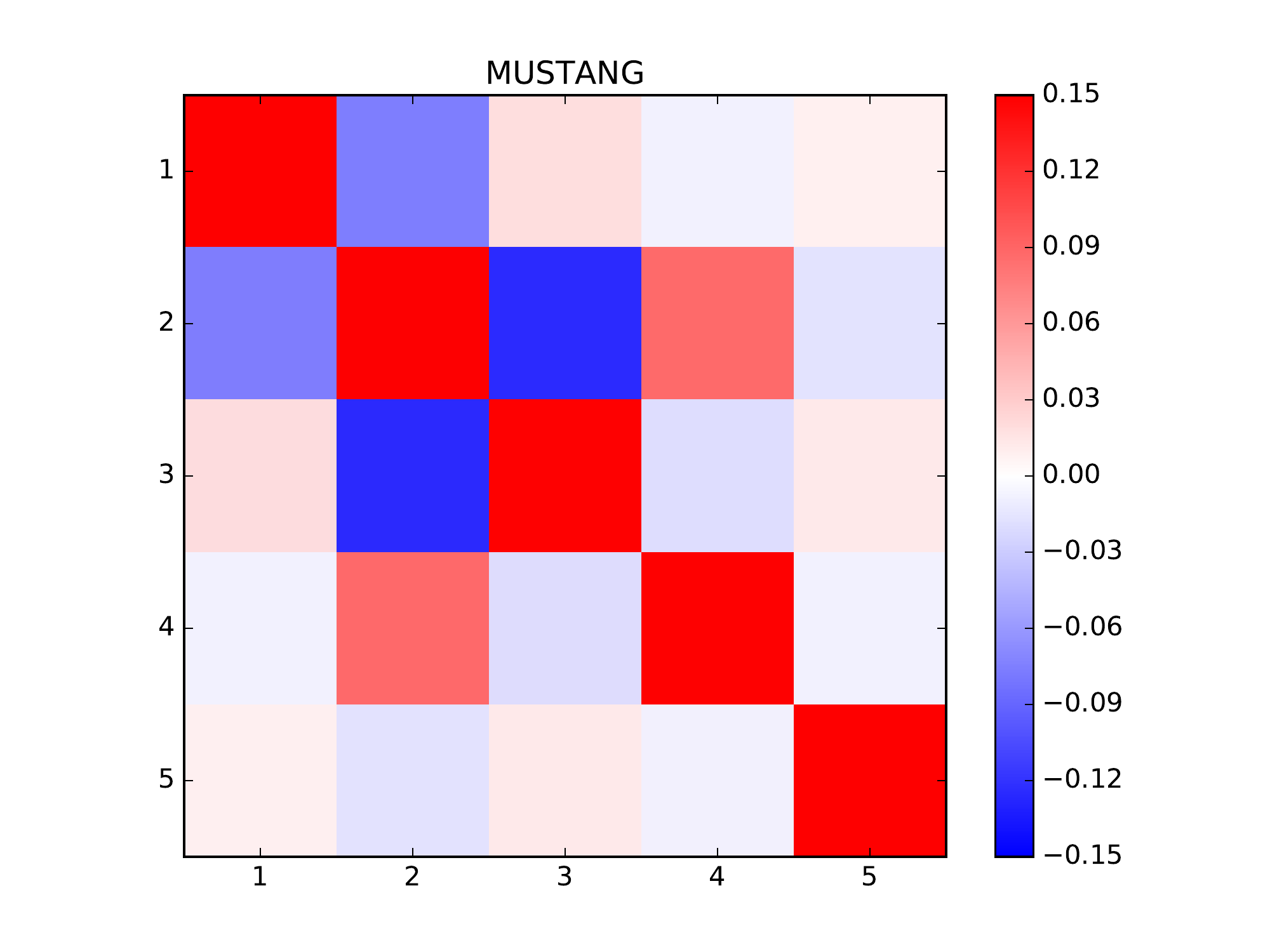}
  \includegraphics[width=0.33\textwidth]{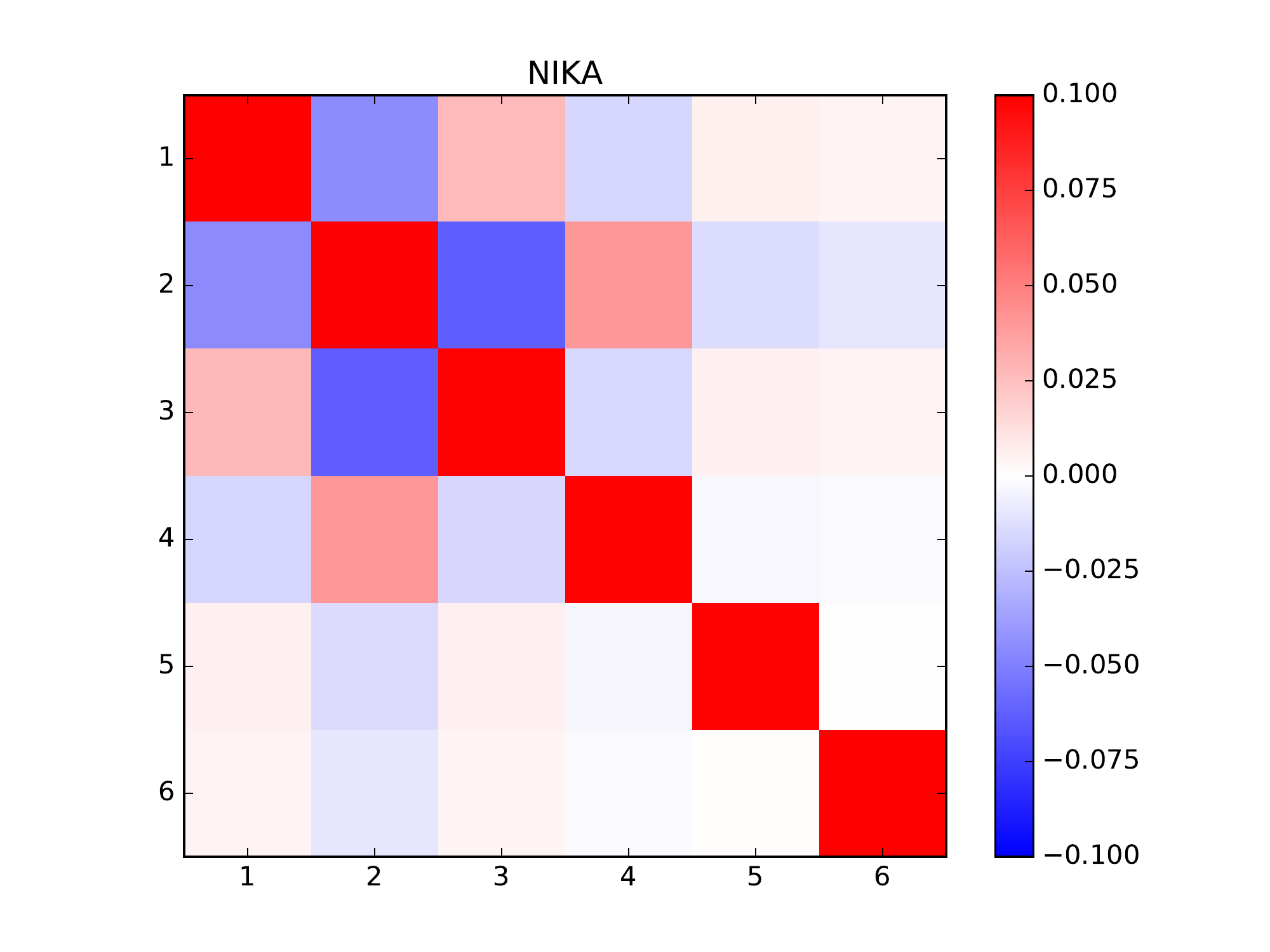}
  \includegraphics[width=0.33\textwidth]{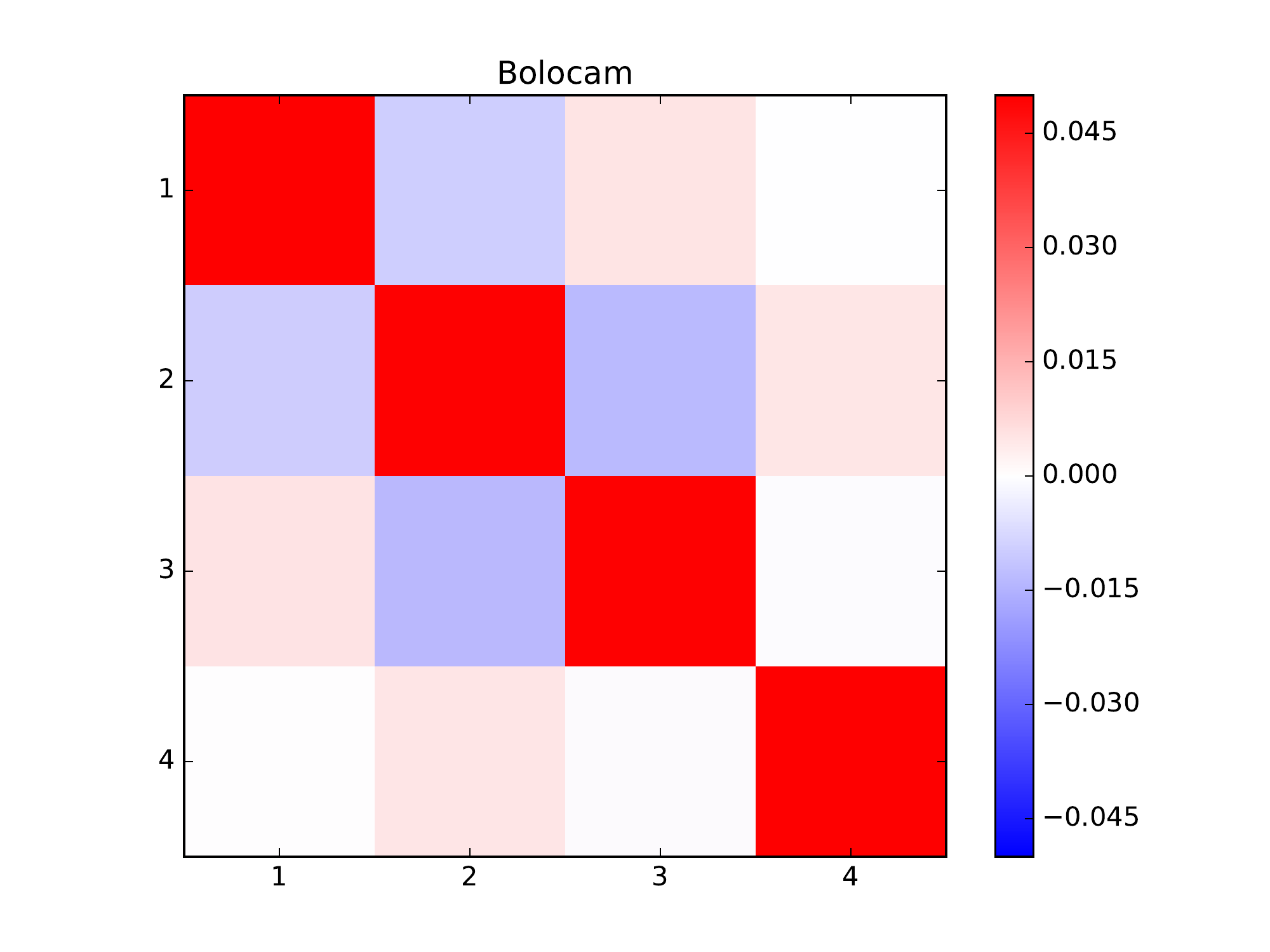}
  \caption{Non-parametric bin correlation matrices. Left: MUSTANG. Middle: NIKA. Right: Bolocam. The coloring is
    scaled to make the magnitude of off-diagonal terms more apparent, and the range changes for each instrument.}
  \label{fig:corr_matrices}
\end{figure*}

We show the correlation matrices in Figure~\ref{fig:corr_matrices}. We notice
that any two adjacent bins are negatively correlated, and by extension, bins spaced 2 apart (e.g bins 1 and 3) are positively
correlated. The maximum amplitude of off-diagonal correlations is 0.13, 0.05, and 0.05 for MUSTANG, NIKA, and Bolocam respectively.
These are relatively small correlations, especially when contrasted with \citet{sayers2013}, for example.

\section{Parametric Pressure Profile (gNFW) Fits}
\label{sec:parfits}

We wish to compare our non-parametric fits to each other (testing consistency between instruments)
and to previous results of cluster pressure profiles. We note that at a $z = 0.89$, CLJ 1227 is at a high redshift,
and although only being one cluster, serves as an initial test of the universality of so-called universal pressure
profile \citep{arnaud2010}, which was derived from a local ($z < 0.2$) sample of clusters.
Given the prevalence of parametric pressure profiles in previous analyses, and in particular, the gNFW parameterization,
we fit a gNFW profile to our non-parametric pressure profile constraints. The gNFW profile is given as:
\begin{equation}
  \Tilde{P} = \frac{P_0}{(C_{500} X)^{\gamma} [1 + (C_{500} X)^{\alpha}]^{(\beta - \gamma)/\alpha}}
  \label{eqn:gnfw_profile}
\end{equation}
where $X = R / R_{500}$, and $C_{500}$ is the concentration parameter; one can also write ($C_{500} X$) as
($R / R_p$), where $R_p = R_{500}/C_{500}$. The exponentials $\alpha$, $\beta$, and $\gamma$ are commonly
cited as the (logarithmic) slopes at moderate, large, and small radii. However, $\alpha$ should be
understood to govern the rate of turnover between the two slopes, $\beta$ and $\gamma$. 


We aim to constrain all parameters within the gNFW profile, but find that $\alpha$ is driven to high values, and
furthermore the constraints are very poor for these high values. Therefore, we choose to restrict $\alpha$ to 1.05,
the value found in \citetalias{arnaud2010}. We further include nuisance parameters of calibration offsets for each dataset.
The calibration uncertainties for NIKA, MUSTANG, and Bolocam are taken to be 7\%, 10\%, and 5\% respectively.
The mean level in each dataset has already been removed or fitted, so it is not considered here. We use the full
covariance matrices from our non-parametric fits.


\subsection{Parametric Constraints}
\label{sec:parametric}

\begin{figure}[!h]
  \centering
  \includegraphics[width=0.5\textwidth]{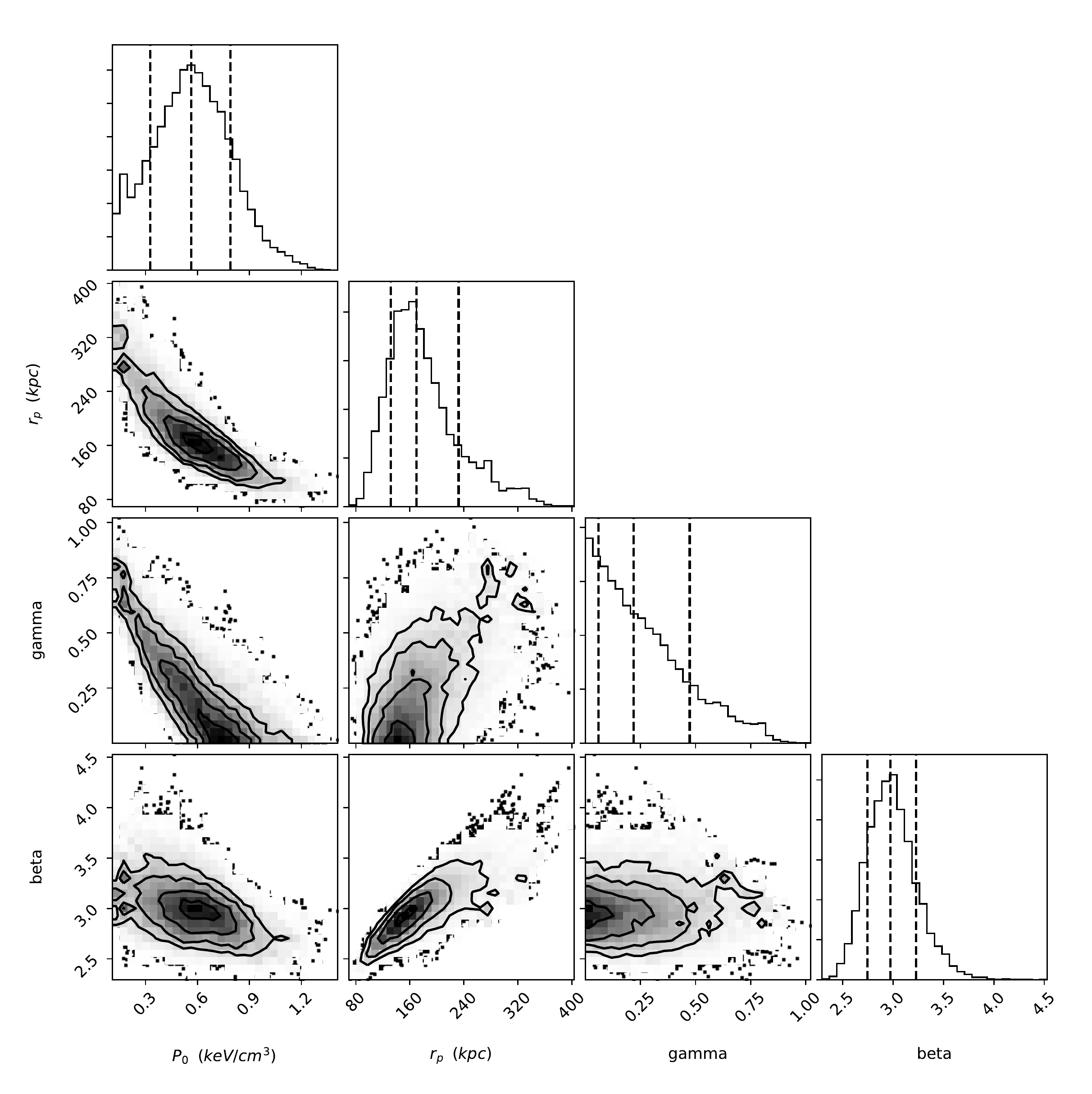}
  \caption{Parameter constraints for our gNFW model: $P_0$, $r_p$, $\gamma$, and $\beta$. Recall that $C_{500} = R_{500} / R_p$.}
  \label{fig:joint_constraints}
\end{figure}
\begin{figure}[!h]
  \centering
  \includegraphics[width=0.5\textwidth]{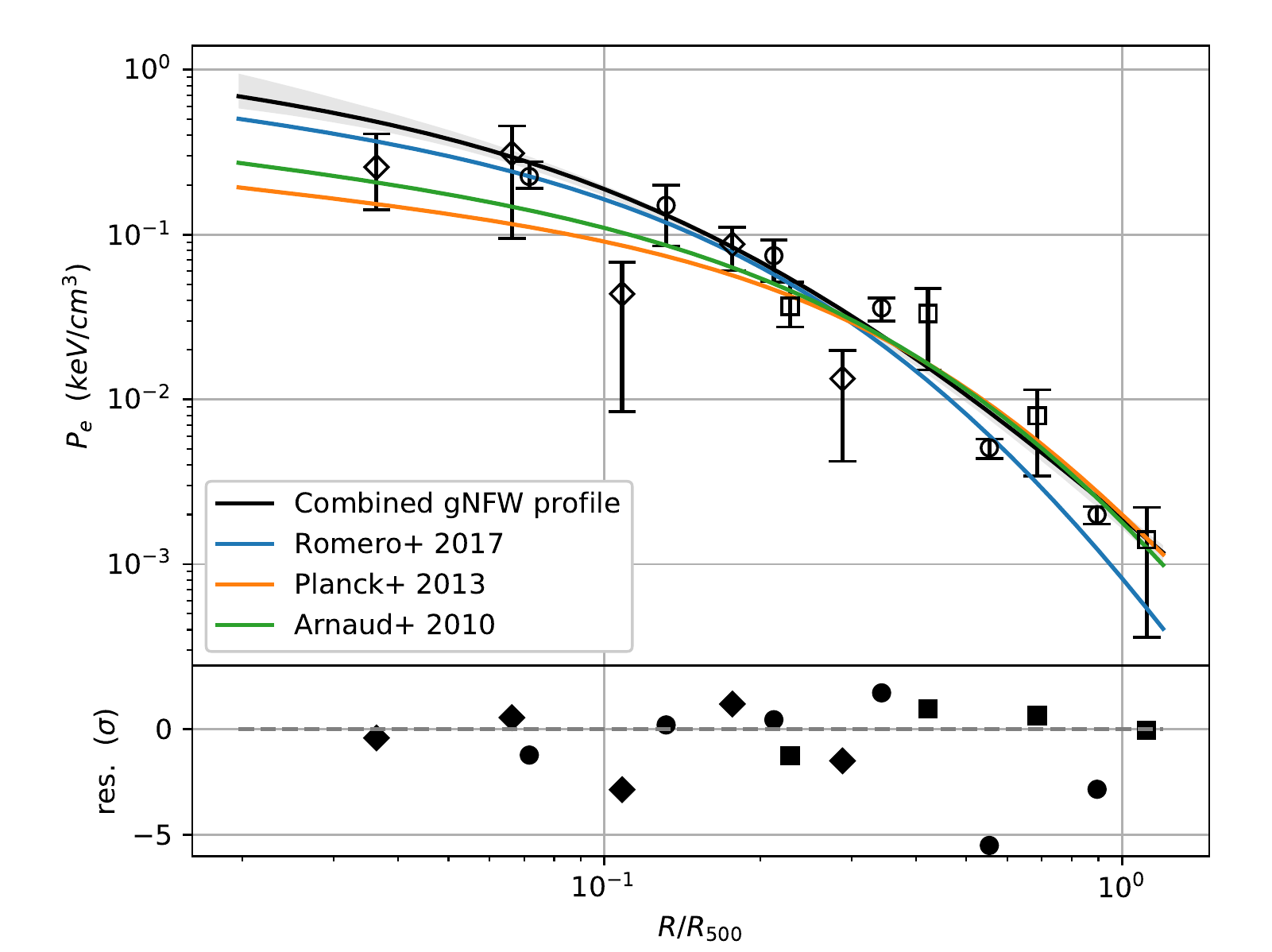}
  \caption{The gNFW (parametric) pressure profile as simultaneously fit to the non-parametric pressure profiles.
    The error bars are statistical, from the MCMC fits. The residual significances ($\sigma$), accounting for calibration errors, are plotted
    in the lower panel.
    The non-parametric symbol-to-instrument association remains the same as in Figures~\ref{fig:virt_robustness} and \ref{fig:nppps}.
    The MUSTANG point that falls well ($\sim-2.8\sigma$) below the gNFW pressure profile (close to 0.1 $R_{500}$) is of
    note and discussed in Section~\ref{sec:discussion}. The last two NIKA points fall $-5.5\sigma$ and $-2.8\sigma$ below the gNFW
    profile.}
  \label{fig:joint_pressure}
\end{figure}


We find gNFW parameters of [$P_0$, $C_{500}$, $\beta$, and $\gamma$] =
[$49.7_{-24.9}^{+22.4}$, $5.89_{-1.78}^{+1.94}$,$2.98_{-0.23}^{+0.28}$, and $0.23_{-0.17}^{+0.30}$].
The power law slope $\gamma$ is within the typical value range found in previous gNFW constraints
\citep[e.g]{nagai2007,arnaud2010,sayers2013},
on CLJ1226 as well as general cluster samples. However, our value of $\beta$ is less than expected;
moreover, for a non-rotating spherical cluster in HSE under thermal pressure support $\beta \le 4$ would
indicate an unbounded mass at arbitrarily large radii (Appendix~\ref{sec:finite_mass}).
Similarly, $P_0$ and $C_{500}$ are larger than generally found.
Given the degeneracy between $\beta$, $P_0$, and $C_{500}$, as shown in Figure~\ref{fig:joint_constraints},
and shape of the pressure profile, these atypical
values of $\beta$ and  $P_0$ appear to be driven by $C_{500}$ being pushed to larger values, where a large $C_{500}$
value indicates that the scale radius (transition in pressure profile slopes) occurs at a relatively small radius.

We also note that the value of $C_{500}$ itself may not be nearly as high if a smaller value of $R_{500}$
is adopted (implying a smaller $M_{500}$ and $P_{500}$.) This may well be the case, as several other
studies conclude that $R_{500} < 1000$ kpc \citep[e.g][]{rumsey2016,mroczkowski2009}.

Setting aside the variations in parameter values themselves, we see in Figure~\ref{fig:joint_pressure} that our
gNFW fit is in agreement for $R \gtrsim 0.3 R_{500}$. In the central regions, our fit shows greater pressure than
would be inferred from \citetalias{arnaud2010} or \citet{planck2013a}, but is consistent with \citet{romero2017}.

\section{Discussion}
\label{sec:discussion}

Our non-parametric fits are well reproduced with varying input parameters (Section~\ref{sec:validity}).
This procedure can be readily applied to ellipsoidal cluster geometries, and could also be modified
to include shock components. Given the potential for ellipsoidal clusters and presence of shocks,
we find that the ability to analyze both the global and local electron pressure in clusters within a
non-parametric approach will be of considerable utility as sensitive, high-resolution,
SZ observations of individual clusters become more commonplace, especially at high redshift.


While our estimation of outer pressure bins may be influenced by the mean level in a map, or a poorly constrained
transfer function, we see that the inner bins remain largely unaffected (Section~\ref{sec:validity}).
We find good agreement in our non-parametric fits between MUSTANG, NIKA, and Bolocam, as all but two points
lie within $2.5\sigma$ of the fitted gNFW profile. The inner point that falls below the gNFW profile comes from
MUSTANG fits, and is only $\sim 2\sigma$ discrepant from the gNFW profile.

This deviation (at a radius of $\sim$12\asec or $0.1 R_{500}$) is consistent with the location of the point source found in
\citet{korngut2011}, and performing a fit on mock observations, where we add a 0.5 mJy source (at 90 GHz)
at this location,
can reproduce the observed deviation. Within the NIKA (150 GHz) data, no evidence for a weak point source is seen,
although, we note that simulated observations of a 1.4 mJy source at the same radial distance does not have a
significant effect on the non-parametric fits, relative to the fits of the simulated observations without a point source.
At other wavelengths, in the 260 GHz NIKA data \citep{adam2015}, as well as at lower frequencies and higher frequencies
(Section~\ref{sec:preprocessing}), no evidence is seen for a point source.


Within our gNFW fits, if $\alpha$ is left unconstrained, we find that large values of $\alpha$ are preferred, indicating a
rapid transition between the inner and outer pressure profile slopes. This turnover is largely driven by NIKA, where, in
Figure~\ref{fig:joint_pressure} (with $\alpha$ fixed), the outer two points fall $-5.5\sigma$ and $-2.8\sigma$ away from
the fitted gNFW pressure profile. NIKA has the best coverage in the spatial region where this transition occurs, and additionally,
NIKA has the strongest detection of the cluster and places the greatest constraints on the pressure profile, globally.


Our gNFW pressure profile fit shows a higher core pressure than that of other sample-averaged gNFW profiles
(see Figure~\ref{fig:joint_pressure}). This is indicative of the cluster being a cool-core cluster \citepalias{arnaud2010}.
However, X-ray data show that the core is relatively hot and is indicative of an ongoing merger \citep[][]{maughan2007}.
With additional support from weak lensing \citep{jee2009a}, and their own AMI data, \citet{rumsey2016}
propose that CLJ 1227 is the early stages of a head-on merger.

This merger scenario could still be consistent with our findings, as a merger affecting the central pressure should indeed
increase the pressure there. Provided that the system is not in equilibrium, and in particular, that increased gas pressure
has not propogated to the outer extent of the cluster, then this scenario would be consistent with the cluster pressure profiles
that we have reconstructed (both the non-parametric and subsequent parametric profiles). An approach of determining non-parametric
profiles, such as that presented here, will be useful for a more accurate analysis of pressure fluctuations and will inform
the degree of non-thermal pressure support \citep[e.g.][]{khatri2016}.



\section{Conclusions}
\label{sec:conclusions}

We developped an algorithm to determine a non-parametric pressure profile for galaxy clusters.
This method is of particular utility to SZ observations, where the filtering effects from data
processing favor fitting forward modelled pressure profiles, as opposed to deriving non-parametric pressure profiles via
geometric deprojection. Our fitting algorithm is robust with respect to input parameters,
bin spacing, and instrumental setup specifics. While the constraints of single-dish SZ observations
beyond the FOV for a given instrument are poor, we find that the inclusion of such a bin
appears to improve the robustness of the pressure constraints within the FOV.

We have applied this algorithm to SZ observations of the high redshift cluster ($z = 0.89$)
CLJ 1227 from MUSTANG, NIKA, and Bolocam. In doing so, we cover a radial range
$0.05 R_{500} < r < 1.1 R_{500}$, continuously recovering spatial scales in this range,
and find consistency among the non-parametric fits of the individual instruments. Furthermore,
parametric best fits indicate a gNFW profile with a relatively small scale radius ($r_p$)
($r_p = C_{500}/R_{500}$). If left unconstrained, $\alpha$ tends towards large values, indicating
a rapid transition at this scale radius between the inner and outer slope. This rapid transition
is consistent across all three instruments, where NIKA is most sensitive to this transition
region and indeed NIKA data alone favors a rapid transition. This rapid transition is also
supported by MUSTANG data, in part due to the drop in recovered pressure at a radius,
$9$\asecs $< r < 23$\asec ($0.07 R_{500} < r < 0.18 R_{500}$).

Empirical investigations into potential point source contamination within this region ($9$\asec $< r < 23$\asec)
indicate that such a point source would have to be $\sim 0.5$ mJy at 90 GHz.
However, the lack of support for such a point source at other wavelengths leads us to doubt this potential explanation for
the dip in MUSTANG pressure between $9$\asecs $< r < 23$\asec.

Our non-parametric fits of the pressure profile of CLJ 1227 are consistent with a smooth (parameterized) pressure profile.
Yet, we have the advantage that deviations from a parameterized pressure profile will be more evident, localized, and
allow for easier investigation of potential contamination or deviations from hydrostatic equilibrium. In its current implementation,
this approach is relatively intuitive, robust, and fast (due to the analytic integration). While a spherical cluster was assumed
for this analysis, the approach already allows for an ellipsoidal geometry. We also foresee the potential to extend this approach
to include analysis of slices within an ellipse, which will prove useful for investigating shocks. We anticipate that this
versatility will be useful in analysis of the NIKA2 SZ large program \citet{comis2016} and other future SZ observations.

\section*{Acknowledgements}

The National Radio Astronomy Observatory is a facility of the National Science Foundation which is operated
under cooperative agreement with Associated Universities, Inc. MUSTANG data was retrieved from
\url{https://safe.nrao.edu/wiki/bin/view/GB/Pennarray/MUSTANG_CLASH}. Original MUSTANG data was
taken under NRAO proposal IDs GBT/09A-052, GBT/09C-059. NIKA data of CLJ 1227 can be found at
\url{http://vizier.cfa.harvard.edu/viz-bin/VizieR?-source=J/A+A/576/A12}. Bolocam data was retrieved from
\url{http://irsa.ipac.caltech.edu/data/Planck/release\_2/ancillary-data/bolocam/}
The Bolocam observations presented here were obtained from the Caltech Submillimeter Observatory, which,
when the data used in this analysis were taken, was operated by the California Institute of Technology under
cooperative agreement with the National Science Foundation. Bolocam was constructed and commissioned using funds
from NSF/AST-9618798, NSF/AST-0098737, NSF/AST-9980846, NSF/AST-0229008, and NSF/AST-0206158. Bolocam observations
were partially supported by the Gordon and Betty Moore Foundation, the Jet Propulsion Laboratory Research and
Technology Development Program, as well as the National Science Council of Taiwan grant NSC100-2112-M-001-008-MY3.

We would like to thank the IRAM staff for their support during the NIKA campaigns. 
The NIKA dilution cryostat has been designed and built at the Institut N\'eel. 
In particular, we acknowledge the crucial contribution of the Cryogenics Group, and 
in particular Gregory Garde, Henri Rodenas, Jean Paul Leggeri, Philippe Camus. 
This work has been partially funded by the Foundation Nanoscience Grenoble, the LabEx FOCUS ANR-11-LABX-0013 and 
the ANR under the contracts ``MKIDS'', ``NIKA'' and ANR-15-CE31-0017. 
This work has benefited from the support of the European Research Council Advanced Grant ORISTARS 
under the European Union's Seventh Framework Programme (Grant Agreement no. 291294).
We acknowledge fundings from the ENIGMASS French LabEx (R. A. and F. R.), 
the CNES post-doctoral fellowship program (R. A.), the CNES doctoral fellowship program (A. R.) and 
the FOCUS French LabEx doctoral fellowship program (A. R.).


\appendix

\section{Analytic Integrals of Ellipsoidally Symmetric Power Laws}
\label{sec:analytic_integrals}

In our non-parametric pressure bin analysis, we assume that the pressure distribution is spherically symmetric.
As the formalism is applicable to ellipsoidally symmetric systems, we present the formulations in ellipsoidal
generality, with the condition that an axis (taken as the $z$ axis) is along the line of sight.
Our quantity to be integrated along the line of sight is denoted as $\epsilon$, and has the
following behavior:
\begin{equation}
  \epsilon(x,y,z) = \epsilon_i (\frac{x^2}{a_i^2}+\frac{y^2}{b_i^2}+\frac{z^2}{c_i^2})^{-P_i},
\end{equation}
where $\epsilon_i$ is a normalization for the pressure within bin $i$;
$a_i$, $b_i$, and $c_i$ are the ellipsoidal scalings of their respective axes,
with the $z$-axis being along the line of sight, and $-2P_i$ is the slope of the pressure profile. We define an
ellipsoidal radius, $r_e = (\frac{x^2}{a_i^2}+\frac{y^2}{b_i^2}+\frac{z^2}{c_i^2})^{1/2}$.
A pressure bin can be in one of three cases: (C1) an ellipsoid of finite extent, (C2) a shell of
finite extent, and (C3) a shell of infinite extent. We use these markers (C1, C2, and C3) as superscripts
when writing definitions per case. The pressure distribution can be rewritten as follows:

\begin{align}
  \epsilon^{\text{C1}}(r_e) &= \begin{cases}
    \epsilon_i \cdot (r_e^2)^{-P} &: r_e^2 \leq 1 \\
                        0 &: r_e^2 > 1,
    \end{cases}
  \label{eqn:case1_shell} \\
  \epsilon^{\text{C2}}(r_e) &= \begin{cases}
                         0 &: r_e^2 < 1 \\
    \epsilon_i \cdot (r_e^2)^{-P} &: 1 \leq r_e^2 \leq R_i^2 \\
                         0 &: r_e^2 > R_i^2, \text{ and}
    \end{cases}
  \label{eqn:case2_shell} \\
  \epsilon^{\text{C3}}(r_e) &= \begin{cases}
                        0 &: r_e^2 < R_i^2 \\
    \epsilon_i \cdot (r_e^2)^{-P} &: R_i^2 \leq r_e^2, 
    \end{cases}
  \label{eqn:case3_shell} 
\end{align}

where $R_i$ is a boundary radius (the outer boundary in Case 2). Given a cluster profile with more than 3 bins, we end
up with many bins in Case 2, in which case we rescale $a_i$, $b_i$, $c_i$, and subsequently $R_i$ each time to properly
normalize each bin ($\epsilon_i$)

Let us define
\begin{equation}
  \kappa = \sqrt{\pi}\epsilon_i c \frac{\Gamma(P_i-0.5)}{\Gamma(P_i)}A^{1-2P_i},
\end{equation}
where $A^2 = (x^2 / a_i^2) + (y^2 / b_i^2)$. While the integration of each bin will share this expression,
the actual values may change depending on $a_i$, $b_i$, and $c_i$ used for
each bin (as above, when multiple bins fall into Case 2). We write the  integration of $\epsilon(r_e)$ along the line of sight as:
\begin{equation}
  I = \int_{-z_0}^{z_0} \epsilon(r_e) dz,
  \label{eqn:los_int_setup}
\end{equation}
where $z_0$ is the outer limit (in $z$) of the region in question. Over the three cases, the solutions are as follows:
\begin{equation}
  I^{\text{C1}} = \left\{
  \begin{array}{lr}
    \kappa (1-I_{A^2}(P_i-0.5,0.5)) &: A^2 \leq 1 \\
    0 &: A^2 > 1
  \end{array}
  \right.
  \label{eqn:case1_int}
\end{equation}
\begin{equation}
  I^{\text{C2}} = \left\{
  \begin{array}{lr}
    \kappa (I_{A^2}(P_i-0.5,0.5)-I_{A^2/R_i^2}(P_i-0.5,0.5)) &: A^2 < 1 \\
    \kappa (1-I_{A^2/R_i^2}(P_i-0.5,0.5)) &: 1 \leq A^2 \leq R_i^2 \\
    0 &: R_i^2 \leq A^2
  \end{array}
  \right.
\end{equation}
\begin{equation}
  I^{\text{C3}} = \left\{
  \begin{array}{lr}
    \kappa (I_{A^2/R_i^2}(P_i-0.5,0.5)) &: A^2 \leq R_i^2 \\
    \kappa &: A^2 > R_i^2
  \end{array}
  \right.
\end{equation}

Here, $I_{A^2}$, or $I_{A^2/R_i^2}$ is the incomplete beta function, often denoted as $I_x(a,b)$. For the discussion of
the gamma and incomplete beta function (below), $x$, $y$, $a$, and $b$ serve as dummy variables. Given our use of
the gamma and incomplete beta functions, it is important to recognize their limitations.
Specifically, $\Gamma(a)$ is undefined for $a = -j, j \in \mathbb{N} \cup \{0\}$ (negative integers, including
zero). The incomplete beta function, having $a = P_i -0.5$ and $b = 0.5$ suffers from undefined values
for $P_i = 0.5-j, j \in \mathbb{N} \cup \{0\}$ as well as $P_i = -j, j \in \mathbb{N} \cup \{0\}$. Finally, all incomplete
beta functions are generally defined for $B(a,b)$ that $Re(a) > 0$ and $Re(b) > 0$. However, the relation of the
incomplete beta function ($I_x$):
\begin{equation}
  I_x(a,b) = I_x(a+1,b) + \frac{x^a (1-x)^b}{a B(a,b)}
  \label{eqn:recibeta}
\end{equation}
allows us to extend the function into the negative domain (for $a$, which we take as $P_i-0.5$). 

To deal with the limitation, generally seen as: $2*y-2 = -j, j \in \mathbb{N} \cup \{0\}$, we derive another approach.
From Equation~\ref{eqn:los_int_setup}, we can substitute variables ($t = z / (cA)$) to arrive at:
\begin{align}
  I &= 2 \epsilon_i A^{-2P_i} \int_{0}^{t_0}(1+t^2)^{-P_i} c A dt \text{ and now adopt } t^2 = \frac{u}{1-u} \\
    &= 2 \epsilon_i A^{-2P_i} \int_{0}^{\theta_0}(1+\tan^2(\theta))^{-P_i} \sec^2(\theta) d\theta \\
    &= 2 \epsilon_i A^{-2P_i} \int_{0}^{\theta_0}\cos^{2P_i-2}(\theta) d\theta
\end{align}

This must then be extended, and is done so with the relation:
\begin{equation}
  \int \cos^{n-2}(\theta) d\theta = \frac{n}{n-1}\int \cos^n(\theta)d\theta - \frac{1}{n-1}\cos^{n-1}(\theta)\sin(\theta) 
  \label{eqn:cosext}
\end{equation}
Given the values of interest/applicability ($2y-2 = -j, j \in \mathbb{N} \cup \{0\}$), this extension is perfectly
applicable, and we will end in well-behaved functions; either:
\begin{align*}
  \int \cos^n(\theta)d\theta &= \tan(\theta) \text{ for } n=-2 \text{ or: } \\
  \int \cos^n(\theta)d\theta &= \ln \vert \sec(\theta) + \tan(\theta) \vert \text{ for } n=-1
\end{align*}

The \textbf{only} case where this analytic integration fails is for $P_i < 0.5$ \textbf{when} integrating out to infinity,
which is fine, as this must diverge in any case. 

\section{Requirements for finite mass for a non-rotating, spherical object under HSE}
\label{sec:finite_mass}

For a non-rotating, spherical object in hydrostatic equilibrium (HSE) under thermal pressure support, we have:
\begin{equation}
  \frac{1}{r^2}\frac{d}{dr}\left( \frac{r^2}{\rho}\frac{dP}{dr} \right) = -4\pi G \rho,
  \label{eqn:landau_hse}
\end{equation}
where $G$ is the Newtonian constant of gravity and $\rho$ is the (total) matter density \citep{landau1959}.
Moreover, we can integrate the first derivative and find:
\begin{equation}
  M(r) = \left( - \frac{dP}{dr} \right) \frac{r^2}{G \rho}
  \label{eqn:mass_hse}
\end{equation}
where we have used the integral
\begin{equation}
  M(R) = \int_{0} ^R (4 \pi \rho r^2 dr),
  \label{eqn:mass_def}
\end{equation}
and note that under this formulation $\rho$ must have a dependence of $r^{-\delta}$, where $\delta > 3$ to have a finite mass
at an arbitrarily large radius. Therefore, returning to Equation~\ref{eqn:mass_hse}, we find that M(r) can be written as:
\begin{equation}
  M(r) =   \left(\alpha P_0 r^{-1-\alpha} \right) \frac{r^2}{G \rho_0 (r^{-\delta})},
\end{equation}
where $P_0$ and $\rho_0$ are simply normalizations of the pressure and density respectively. The mass under hydrostatic
equilibrium will then have the radial dependence as $r^{1+\delta - \alpha}$, where again, $\delta > 3$. Therefore, we find
that $\alpha \ge 4$ is required for a finite mass of an object, under the assumptions stated above.


\bibliography{mycluster}
\label{references}

\end{document}